# scientific reports

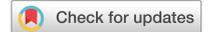

**OPEN**

# Natural quantum reservoir computing for temporal information processing

Yudai Suzuki[1]✉, Qi Gao[2,3], Ken C. Pradel[2], Kenji Yasuoka[1] & Naoki Yamamoto[3,4]

Reservoir computing is a temporal information processing system that exploits artificial or physical dissipative dynamics to learn a dynamical system and generate the target time-series. This paper proposes the use of real superconducting quantum computing devices as the reservoir, where the dissipative property is served by the natural noise added to the quantum bits. The performance of this natural quantum reservoir is demonstrated in a benchmark time-series regression problem and a practical problem classifying different objects based on temporal sensor data. In both cases the proposed reservoir computer shows a higher performance than a linear regression or classification model. The results indicate that a noisy quantum device potentially functions as a reservoir computer, and notably, the quantum noise, which is undesirable in the conventional quantum computation, can be used as a rich computation resource.

Recent remarkable advances in machine learning have attracted more and more attention in the field of bioinformatics[1], computer visions[2,3], finance[4,5], and physics[6,7] due to it offering powerful and efficient techniques for solving various types of tasks in those research areas. One common task is temporal information processing, where sequential or time-series data is processed to achieve a specific goal, such as natural language processing[8] and robotics[9]. To solve temporal information tasks, recurrent neural network (RNN)[10] is often used. The basic strategy to train RNN is to recurrently connect the nodes, so that it approximates a target dynamical system. However, this is a computationally demanding process.

As a special type of RNN, reservoir computing (RC)[11] has been actively studied, which originate from echo state networks (ESNs)[12] and liquid state machines (LSMs)[13]. In RC, the temporal input data is mapped into the state of a high-dimensional dynamical system called the *reservoir*, and then a simple linear regression is used to adjust the weight of the readout signal of the dynamics, allowing for approximation of the target output signal. By virtue of this simple learning process, RC can realize fast and stable learning at a lower computational cost compared to typical RNNs.

The reservoir serves as a nonlinear pattern mapping of the input signal to high-dimensional dynamical states, as do kernel functions in kernel methods[14]. Thus, choosing the right type of reservoir systems to be implemented is of significance. In the last few decades, many different implementations of reservoirs have been proposed, from the original ESN and LSM models to *physical reservoir* such as field programmable gates arrays (FPGAs)[15], a bucket of water[16], soft robotics[17], tensegrity-structured robots[18], and spintronics[19]. Demonstrations of physical reservoir implementation has attracted increased attention, because physical reservoirs could potentially process information faster at a lower computational cost of learning and with higher performance[20,21]. In this direction, quantum systems are proposed as a promising candidate for the physical reservoir[22]. This is because of the common thought corroborated by theoretical and experimental results[23–26] that large quantum systems would be in general hard to simulate by classical computers, and thus the quantum reservoirs (QRs) with intrinsic complex dynamics are conjectured to show higher performance for some temporal information tasks. In fact, quantum reservoir computing (QRC) has been extensively investigated from theoretical analysis of the QR property[27,28], to performance analysis through numerical simulations[29,30], and various other improvements[31–33]. It has also been applied to quantum tasks such as quantum tomography[34,35]. Also, thanks to advances in quantum hardware, the physical implementations of QRs have been demonstrated in the nuclear magnetic resonance (NMR) ensemble systems[36] and superconducting quantum processors[28]. QRC thus represents a promising direction in the field of quantum machine learning[37–41].

[1]Department of Mechanical Engineering, Keio University, Hiyoshi 3-14-1, Kohoku, Yokohama 223-8522, Japan. [2]Mitsubishi Chemical Corporation, Science & Innovation Center, 1000, Kamoshida-cho, Aoba-ku, Yokohama 227-8502, Japan. [3]Quantum Computing Center, Keio University, Hiyoshi 3-14-1, Kohoku, Yokohama 223-8522, Japan. [4]Department of Applied Physics and Physico-Informatics, Keio University, Hiyoshi 3-14-1, Kohoku, Yokohama 223- 8522, Japan. ✉email: yudai.suzuki.sh@gmail.com

In this paper, we propose a QRC framework that exploits a *natural* quantum dynamics on a gate-based superconducting quantum processors as a reservoir, for temporal information processing. This is in stark contrast to the previously developed *artificial* QRC architecture, where the quantum dynamical system at each timestep is realized by injecting the inputs into one ancillary qubit and driving the whole QR system via an input-independent Hamiltonian, and finally discarding and resetting the ancillary qubit for the next step. Discarding the ancillary qubit in this process is crucial to realize the dissipative nature of the system, which is an essential property for reservoirs. On the other hand, in our QR systems, we instead utilize the intrinsic noise of the quantum devices to realize a dissipative quantum dynamics. That is, we rather positively take advantage of currently available noisy intermediate-scale quantum (NISQ) devices[42] which have unavoidable dissipative noise (decoherence). The idea behind our proposal comes from harnessing this seemingly undesired behavior of dynamical systems as a rich computational resource for the RC framework, such as a soft robot reservoir[17]. Note that in this manuscript, we use the term *natural reservoir* to denote a physical reservoir that is difficult to describe mathematically such as the dynamics of soft robotics in[17], while the term *artificial reservoir* is used to denote those whose mathematical model is in principle available.

The proposed natural QRC approach is a hardware-specific one, and its performance must be experimentally evaluated on a real device. In this work we utilized IBM superconducting quantum processors to demonstrate our QR systems for two temporal information tasks, emulation of the Nonlinear Auto-Regressive Moving Average dynamics (NARMA task) and classification of different objects based on the sensor-data obtained by grabbing them (object classification task). For NARMA task, we observed that our QR systems realized on 'ibmq_16_melbourne' (we simply call the Melbourne device) show higher performance than a linear regression model. As for the object classification task, our QR system on 'ibmq_toronto' (the Toronto device) shows higher classification accuracy than a simple linear classifier for the task of classifying three different objects. These results indicate that a natural QR realized on a noisy quantum device can be considered a promising candidates as a reservoir for the RC framework. At the same time, this work paves the way for using NISQ devices to solve practical problems.

## Methods

**General framework of reservoir computing.** RC utilizes a dynamical system to execute temporal information processing tasks. Typically, the goal of this task is to learn a function that transforms an input sequence (time series) to a target output sequence, for the purpose of time series forecasting[43] and pattern classification[44]. In this work, we consider a given pair of scalar input $\{u_t\}_{t=1}^{M}$ and target output $\{y_t\}_{t=1}^{M}$ signals. The RC model is described as follows;

$$x_t = f(x_{t-1}, u_t), \tag{1}$$

$$\bar{y}_t = W_{out}^T h(x_t), \tag{2}$$

where $x_t$ is the state vector at time $t$ and $\bar{y}_t$ is the corresponding scalar output. $W_{out}$ is the vector of adjustable parameters, which is to be optimized by a learning algorithm, so that the output $\bar{y}_t$ approximates to the target output $y_t$. The function $f$ represents the time evolution of the reservoir state $x_t$, driven by the input $u_t$, and $h$ is a function that observes signals from $x_t$. The point of this RC framework is that the reservoir part is fixed unlike an RNN, meaning that the learning algorithm is much simpler than an RNN. On the other hand, the crucial role of feature extraction is left to the reservoir dynamics, so the choice/design of reservoir is very important.

In the RC framework, the parameter $W_{out}$ is tuned in a supervised manner. The mean squared error (MSE) between the output $\bar{y}_t$ and the target output $y_t$ during the training phase, $\text{MSE} = \sum_{t=t_f}^{t_l} (\bar{y}_t - y_t)^2$, is minimized, where $t_f$ and $t_l$ are the first and last timestep of the training phase, respectively. This can be readily solved via the linear regression technique, given by the pseudo inverse of the linear equation

$$\mathbf{y} = W_{out}^T \mathbf{X}, \tag{3}$$

where $\mathbf{X} = \left(\tilde{h}(x_{t_f}), \ldots, \tilde{h}(x_{t_l})\right)$ and $\mathbf{y} = (\bar{y}_{t_f}, \ldots, \bar{y}_{t_l})$. Here $\tilde{h}(\cdot)$ is the vector containing both $h(\cdot)$ and the bias term 1, i.e. $\tilde{h}(\cdot) = \left(h(\cdot)^T, 1\right)^T$.

**Quantum reservoir computing model.** In the QRC framework, the dynamical system (1) is given by

$$\rho_t = \mathcal{T}_{u_t}(\rho_{t-1}), \tag{4}$$

where $\rho_t$ is the density operator that represents a state of the QR at time $t$, and $\mathcal{T}_{u_t}$ is an input-dependent completely positive and trace preserving (CPTP) map that describes the time evolution of the QR system. The CPTP map considered in Ref.[22] is given by

$$\begin{aligned}\rho_t &= \mathcal{T}_{u_t}(\rho_{t-1}) \\ &= e^{-iH\tau}\left(\rho_{\text{input}} \otimes \text{Tr}_{\text{input}}(\rho_{t-1})\right)e^{iH\tau},\end{aligned} \tag{5}$$

where $\rho_{\text{input}} = |\psi_{u_t}\rangle\langle\psi_{u_t}|$ with $|\psi_{u_t}\rangle = \sqrt{1-u_t}|0\rangle + \sqrt{u_t}|1\rangle$. Here $|0\rangle$ and $|1\rangle$ represent $[1,0]^T$ and $[0,1]^T$, respectively, and $\langle\cdot|$ is the Hermitian conjugate of $|\cdot\rangle$. That is, the input $u_t \in [0,1]$ is injected onto one ancillary qubit, and then the whole QR system is time-evolved by the input-independent unitary operator $e^{-iH\tau}$ with user-defined

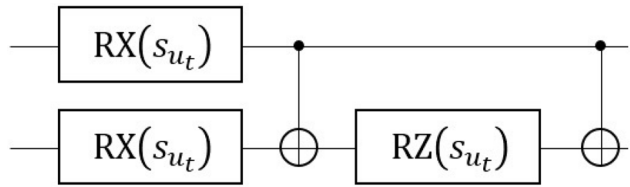

**Figure 1.** Quantum circuit representation of the local 2 qubits unitary operator (9).

Hamiltonian $H$ and time interval $\tau$. The partial trace over the ancillary qubit, represented by $\text{Tr}_{\text{input}}$, is essential to determine the dissipative property of the QRC.

The function $h$ in Eq. (2) is usually given by a set of expectation values of local observables defined by the QR system. For an $n$-qubit QR system, we can take a set of the expectation values

$$h(\rho_t) = [\text{Tr}(Z_1 \rho_t), \ldots, \text{Tr}(Z_n \rho_t)]^{\text{T}}, \tag{6}$$

where $Z_i$ is the Pauli $Z$ matrix on the $i$th qubit, i.e.,

$$Z_i = I \otimes \cdots \otimes Z \otimes \cdots \otimes I, \quad Z = |0\rangle\langle 0| - |1\rangle\langle 1| = \begin{bmatrix} 1 & 0 \\ 0 & -1 \end{bmatrix}, \quad I = \begin{bmatrix} 1 & 0 \\ 0 & 1 \end{bmatrix}.$$

**Natural quantum reservoir.** Here we describe our QR system that, unlike the artificial mixing process $\text{Tr}_{\text{input}}$ given in Eq. (5), makes full use of natural noise as a computational resource for the QRC paradigm.

As mentioned in "Introduction", currently available quantum devices all suffer from quantum noise. However, in the QRC framework, that unwanted quantum noise could be rather beneficial. For example, the QR system with depolarizing error can satisfy one of the essential properties of the reservoir, the echo-state[45] or the convergence property[46], which is in a broad sense a condition where the system asymptotically forgets its initial state. To be more specific, the depolarizing error is given by the following CPTP map:

$$\mathcal{E}_{\text{dep}}(\rho) = (1-p)\rho + p\frac{I}{d},$$

where $\rho$ is a system density matrix with the size of $d$ and $p$ is the parameter proportional to the probability that the quantum state becomes the maximally-mixed state[47]. The convergence property of the QR in this case can be easily checked; if the depolarizing channel is successively applied to the system, $\rho$ asymptotically converges to the maximally-mixed state, $I/d$. In general, the dynamical system under unital error process converges to a stable equilibrium state[48,49]. In addition, the QR system implemented on a noisy device potentially has a desirable "memory" property. It was reported in[50] that a noisy quantum device has some memory effects such as temporal correlation. For these reasons, an open quantum system under natural error is potentially usable as a QR.

Based on the above discussion, we consider the following dynamics as the QR system (4):

$$\begin{aligned} \rho_t &= \mathcal{T}_{u_t}(\rho_{t-1}) \\ &= \mathcal{E}_{\text{device}}\left(U(u_t)\rho_{t-1}U(u_t)^\dagger\right), \end{aligned} \tag{7}$$

where $U(u_t)$ is an input-dependent unitary operator and $\mathcal{E}_{\text{device}}$ is the un-modeled CPTP map corresponding to the real device during operation. In this work, we consider the $n$-qubits system driven by the unitary operator

$$U(u_t) = \bar{U}_{0,1}(u_t) \otimes \bar{U}_{2,3}(u_t) \otimes \cdots \otimes \bar{U}_{n-2,n-1}(u_t), \tag{8}$$

where $\bar{U}_{i,j}(u_t)$ is the identical local 2-qubits unitary operator acting on the $i$th and $j$th qubits:

$$\bar{U}_{i,j}(u_t) = \text{CX}_{i,j}\text{RZ}_j(s_{u_t})\text{CX}_{i,j}\text{RX}_i(s_{u_t})\text{RX}_j(s_{u_t}). \tag{9}$$

Here $s_{u_t} = au_t$ with $a \in \mathbb{R}$ and $\text{CX}_{i,j}$ is the CNOT gate with control qubit $i$ and target qubit $j$. Also $\text{RZ}_i(s) = \exp(-isZ/2)$ and $\text{RX}_i(s) = \exp(-isX/2)$ are the rotation gate applied on the $i$th qubit, around the Pauli $Z$ and $X$ axis, respectively. At $t = 0$, the state is prepared as $|+\rangle = |+\rangle^{\otimes n} = H^{\otimes n}|0\rangle^{\otimes n}$ with $H$ the Hadamard gate. The quantum circuit representation of the local unitary operator $\bar{U}_{i,j}(u_t)$ is depicted in Fig. 1. This 2-qubits unitary operator is a type of hardware-efficient ansatz[39]; as seen from the figure, this unitary can realize a limited transformation of state. The purpose of choosing such a specific unitary gate is to see if the quantum noise could enrich the complexity of the dynamical system. That is, we restricted the type of gate to exclude the possibility that the performance improvement comes from the richness of the unitary gate and clearly see the net effect of natural noise. Note that, in the noiseless situation, Eq. (8) is a trivial dynamics composed of identical and independent 2-qubits subsystems. However, the subsystems implemented on a noisy real device may be able to couple with the neighboring subsystems due to the natural noise introduced from the surrounding environment, such as the crosstalk[51,52], which as a result may lead to non-trivial QR dynamics. Lastly, the output signal at timestep $t$ given by Eq. (6) is obtained by repeatedly acting the quantum circuit on the initial state $\rho_0 = |+\rangle\langle +|$ as

$$\rho_t = \mathcal{T}_{u_t} \circ \mathcal{T}_{u_{t-1}} \circ \cdots \circ \mathcal{T}_{u_1}(\rho_0) \tag{10}$$

and then measuring $\rho_t$ in the computational basis, as shown in Fig. 2. In the experiment, we took 8192 measurement shots for each $t$ to calculate the mean $\mathrm{Tr}(Z_i \rho_t)$ in Eq. (6).

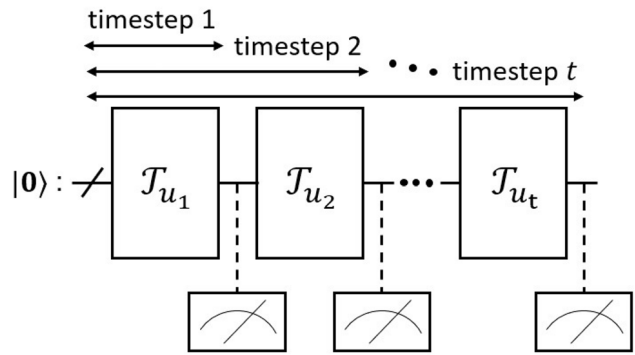

**Figure 2.** Circuit diagram of the proposed QR system. $|0\rangle$ stands for the initial state, and $\mathcal{T}_{u_i}$ is a fixed CPTP map with input $u_i$. The output signal (6) is obtained by repeatedly running the quantum circuit according to Eq. (10) and measuring $Z$ for each qubit.

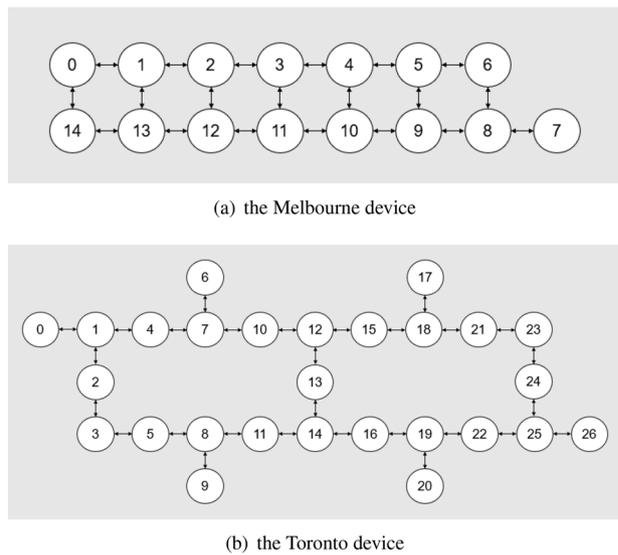

**Figure 3.** Configuration of the NISQ devices used for the experiments, (**a**) the Melbourne device and (**b**) the Toronto device. In the figures, the nodes and edges represent qubits and the physical coupling of qubits, respectively, where the numbers on the nodes are just labels.

### Results and discussion
Here we show the result of our test of the performance of our QRC scheme. We consider two temporal information processing tasks, regression of the NARMA dynamics (NARMA task) and classification of different objects based on the sensor-data obtained by grabbing them (object classification task). We use two IBM superconducting quantum processors; Melbourne and Toronto devices, whose configurations are illustrated in Fig. 3. Also, in the Qiskit package that we used for the experiments to work with the IBM quantum processors, users can determine the "optimization level" of the quantum circuit transpiler for reducing noise caused by sources such as redundant gate operations. In the QRC scenario, we are rather interested in introducing the noise, hence an optimization level of zero was chosen (i.e., no noise reduction).

**NARMA task.** We first demonstrate the performance of our QR systems with the NARMA task. The NARMA task is a benchmark test used to evaluate the performance of dynamical models for temporal information processing, in terms of nonlinearity and memory (dependence on past output) properties[53,54]. The goal of the task is to emulate the dynamics generating the NARMA output sequence $\{y_t\}_{t=1}^{M}$. An example studied in[22,55] is described as

$$y_{t+1} = 0.4y_t + 0.4y_t y_{t-1} + 0.6u_t^3 + 0.1, \qquad (11)$$

where $u_t$ and $y_t$ are the input and target output sequences, respectively. Another NARMA dynamics studied in[22,33,55] is expressed as

$$y_{t+1} = \alpha y_t + \beta y_t \left( \sum_{j=0}^{n_o-1} y_{t-j} \right) + \gamma u_{t-n_o+1} u_t + \delta, \qquad (12)$$

where $(\alpha, \beta, \gamma, \delta) = (0.3, 0.05, 1.5, 0.1)$ and $n_o$ is the order that determines the degree of the nonlinearity. In our experiment, we consider the following three NARMA dynamics; NARMA of Eq. (11) (we call it NARMA2), and NARMAs of Eq. (12) with $n_o = 5$ and $n_o = 10$ (we call them NARMA5 and NARMA10, respectively). Note that the number in the task name (e.g., 5 in "NARMA5") implies the order of the nonlinearlity, and hence these three NARMA tasks can be used to evaluate the nonlinearity of the QR systems.

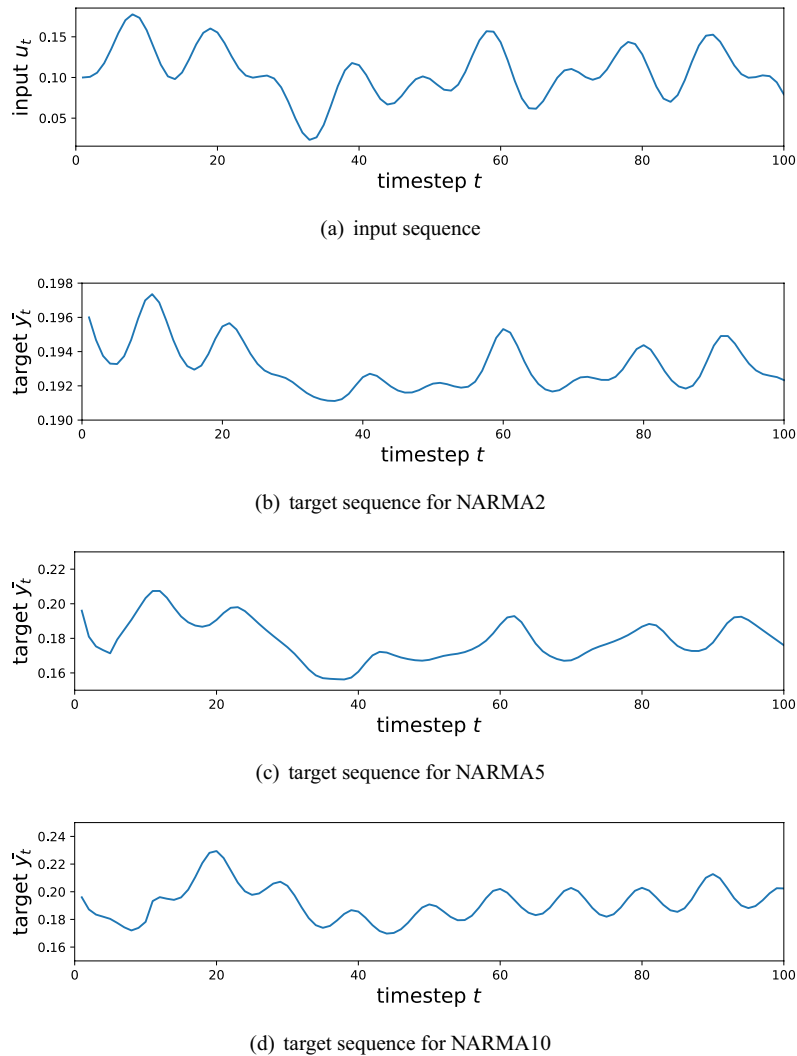

**Figure 4.** Time series data used for the experiment. (**a**) Shows the input sequence. (**b–d**) Shows the target sequences for NARMA2, NARMA5, and NARMA10, respectively.

As for the input $\{u_t\}_{t=1}^M$, we use the following time series for all the NARMA tasks;

$$u_t = 0.1 \left( \sin\left(\frac{2\pi \bar{\alpha} t}{T}\right) \sin\left(\frac{2\pi \bar{\beta} t}{T}\right) \sin\left(\frac{2\pi \bar{\gamma} t}{T}\right) + 1 \right), \qquad (13)$$

where $(\bar{\alpha}, \bar{\beta}, \bar{\gamma}, T) = (2.11, 3.73, 4.11, 100)$, as used in[22]. Here, we set the length of inputs and outputs to $M = 100$, where the first 10 timesteps are used for washout, the following 70 timesteps are used for training, and the remaining 20 timesteps are used for testing. The washout period is necessary for the QR system to lose its dependence on the initial state $\rho_0$, and thus the signals in this period are not used for training. Figure 4 shows the inputs and the target output sequences for each NARMA task.

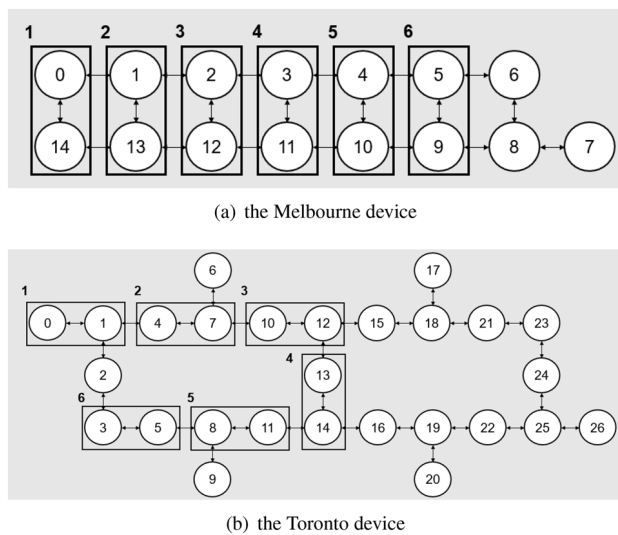

**Figure 5.** Qubits that constitute the subsystems in (**a**) Melbourne device and (**b**) Toronto device. The black box indicates the subsystem with labels from 1 to 6. When we mention "$m$ subsystems are used", this means that the subsystems labeled 1 to $m$ are used.

|  | Quantum model | | | | | | Classical model |
|---|---|---|---|---|---|---|---|
|  | Melbourne device | | | Toronto device | | | |
|  | 4-Subsystems | 5-Subsystems | 6-Subsystems | 4-Subsystems | 5-Subsystems | 6-Subsystems | LR |
| **(a) NARMA2** | | | | | | | |
| Mean | $1.3 \times 10^{-5}$ | $1.3 \times 10^{-5}$ | **$8.9 \times 10^{-6}$** | $2.9 \times 10^{-5}$ | $2.5 \times 10^{-5}$ | $2.2 \times 10^{-5}$ | $1.8 \times 10^{-5}$ |
| Std | $6.3 \times 10^{-5}$ | $2.8 \times 10^{-6}$ | $2.8 \times 10^{-6}$ | $6.7 \times 10^{-6}$ | $1.3 \times 10^{-5}$ | $4.1 \times 10^{-6}$ | — |
| **(b) NARMA5** | | | | | | | |
| Mean | **$1.3 \times 10^{-3}$** | **$1.3 \times 10^{-3}$** | **$1.3 \times 10^{-3}$** | $2.7 \times 10^{-3}$ | $2.2 \times 10^{-3}$ | $1.9 \times 10^{-3}$ | $2.6 \times 10^{-3}$ |
| Std | $6.7 \times 10^{-4}$ | $4.0 \times 10^{-4}$ | $5.3 \times 10^{-4}$ | $9.6 \times 10^{-4}$ | $2.1 \times 10^{-4}$ | $3.7 \times 10^{-4}$ | — |
| **(c) NARMA10** | | | | | | | |
| Mean | $1.9 \times 10^{-3}$ | $2.1 \times 10^{-3}$ | $2.0 \times 10^{-3}$ | $3.6 \times 10^{-3}$ | $3.1 \times 10^{-3}$ | $2.3 \times 10^{-3}$ | **$9.7 \times 10^{-4}$** |
| Std | $4.8 \times 10^{-4}$ | $6.0 \times 10^{-4}$ | $3.5 \times 10^{-4}$ | $8.5 \times 10^{-4}$ | $1.2 \times 10^{-3}$ | $5.3 \times 10^{-4}$ | — |

**Table 1.** List of NMSEs for (a) NARMA2, (b) NARMA5, and (c) NARMA10. For comparison, NMSEs of the classical linear regression model (denoted as LR) are shown. The bold script indicates the best NMSE for each NARMA task.

We use the Melbourne and Toronto devices to run the unitary operator given in Eqs. (8) and (9) with $a = 2$. These devices are used in the same condition to compare the effect of the hardware-specific noises. In particular, we investigate if increasing the system size (accordingly the number of output signals) may improve the performance of QRC, which was numerically predicted in[22]. For this purpose, we study the cases where the number of qubits is chosen as $n = 8, 10,$ and $12$. This means that the number of 2-qubit subsystems is $m = 4, 5,$ and $6$, respectively. The qubits used are shown in Fig. 5.

To quantitatively evaluate the performance of the QRC, we calculate the normalized mean squared errors (NMSE) between the output (2), $\bar{y}_t = W_{out} h(x_t)$, and the target output $y_t$ for the testing period, which is expressed as

$$NMSE = \frac{\sum_{t=81}^{100} (\bar{y}_t - y_t)^2}{\sum_{t=81}^{100} y_t^2}. \tag{14}$$

Table 1 summarizes the NMSE for each experimental setting, where the NMSEs are averaged over 10 experimental trials under the same conditions. The experiments had been performed during the period between Aug. 16th and Nov. 2nd in 2020. To see the effect of nonlinearity induced by the QR systems, we compared our model with a simple linear regression (LR) model that predicts the output by $\bar{y}_{t+1} = w u_t + b_0$ with optimized parameters $w$ and $b_0$. Also Figs. 11, 12 and 13 shows the result for each NARMA task. Below we list the summary of the results, depending on the type of task.

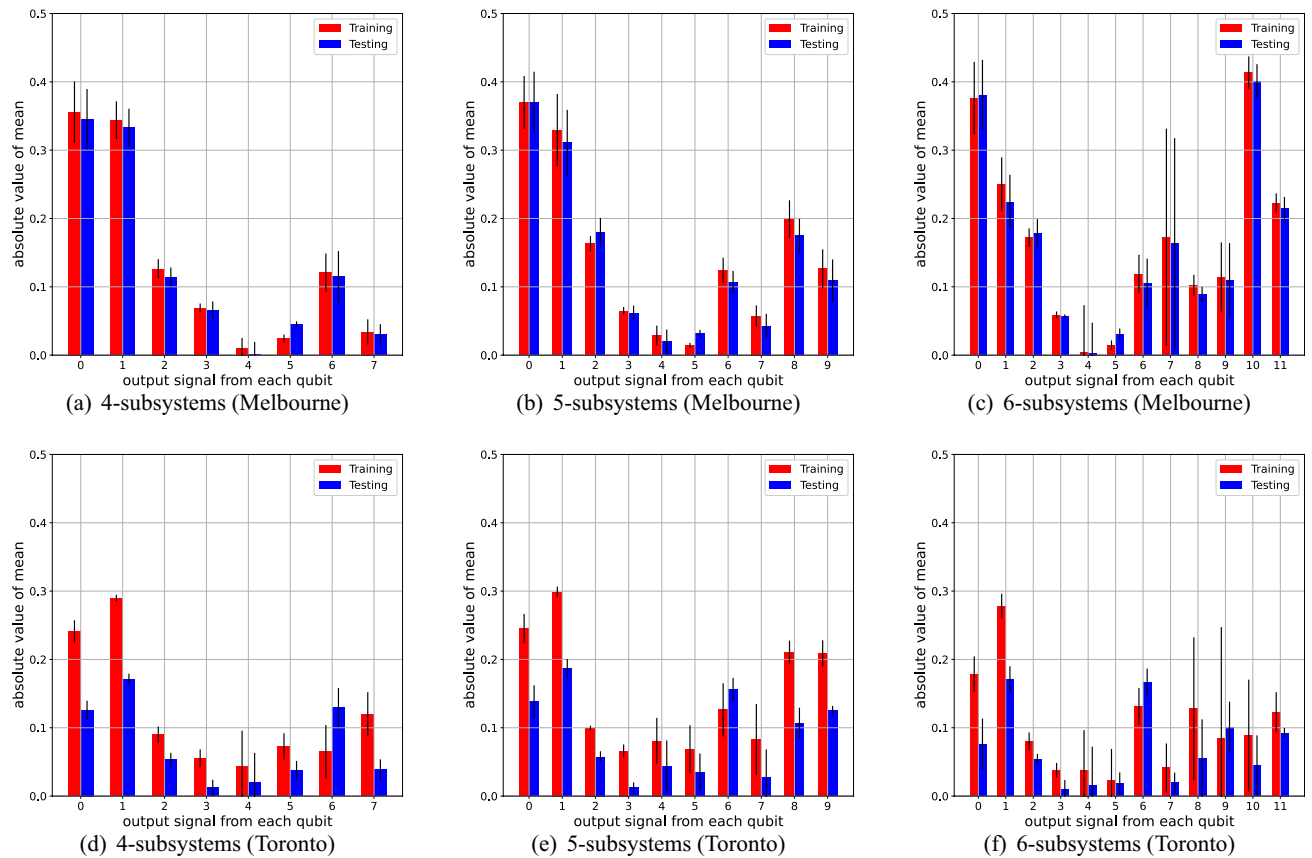

**Figure 6.** Mean of the output signal from each qubit over training (red) and testing (blue) phases. Note that we show the absolute value of the mean for the sake of the convenience. The number on the horizontal axis indicates the label of the qubit from which the signal is measured. The upper three panels (**a**–**c**) and lower three panels (**d**–**f**) show the results for the Melbourne and Toronto device, respectively.

- For the NARMA2 task, all the QR systems on the Melbourne device outperformed the LR model. In particular, a larger QR system showed higher performance with the QR with 6 subsystems showing the lowest NMSE among all models. On the other hand, every QR system on the Toronto device was inferior to the LR model.
- We found a similar tendency for the NARMA5 task. All the QR systems on the Melbourne device were better than the LR model. However, the performance of the QR system did not improve even by increasing the system size in this task. On the other hand, for the QR systems on the Toronto device, the performances of QRs with 5 or 6 subsystems was higher than the LR model. Also the larger system shows a better performance.
- As for the NARMA10 task, the QR systems on both Melbourne and Toronto devices could not outperform the LR model. Also, the tendency with respect to the system size differed from the cases of the NARMA2 and NARMA5. The performance improved when the system size of the Toronto device was increased, while that of the Melbourne device did not improve.

These results show that the performance of our QRC scheme heavily depends on the device; the Melbourne device is always better than the Toronto device for the tasks with the same experimental setting. Also, we find the different tendency of these devices with respect to the number of subsystems. Namely, the performance of the Toronto device is monotonically improved by increasing the system size for all the NARMA tasks, while the Melbourne device does not show this tendency, except for the NARMA2 task.

These device-dependent features can be partially explained by analyzing the output signals from the QR systems. In the analysis, we focus on the the *stationarity* of the output signals. Broadly speaking, stationarity is a notion that represents the time-consistent property of time-series data. Here, to see the stationarity of output signals from the devices, we calculate their mean and variance over the training and testing phase, which are summarized in Figs. 6 and 7, respectively. From Fig. 6, we observe that, for the case of the Melbourne device, the mean values of the output signals over the training and testing phase are relatively close for every qubit, meaning that the output signals are stationary in view of the mean. On the other hand, some qubits on the Toronto device (say, qubits labeled 0 and 1) witnessed a large gap in the mean value between the training and test phase, implying that the output signals are not stationary. Further, the non-stationarity of the Toronto device can also be seen in view of the variance. In Fig. 7, although both devices experience a decrease in variance when changing the phase, the gap for the Toronto device is more significant. (Note that a logarithmic scale is used for the vertical axis of the plots in Fig. 7.) Consequently, for the Toronto device, the output signals used for training vary greatly from those in the testing phase, which might be one reason why the Toronto device is inferior to the Melbourne. Also, because the target output for every NARMA task is stationary in terms of both the mean and the variance, as shown in Table 2, the non-stationary dynamics of the Toronto device cannot track the target well, which is also one reason of the inferior performance.

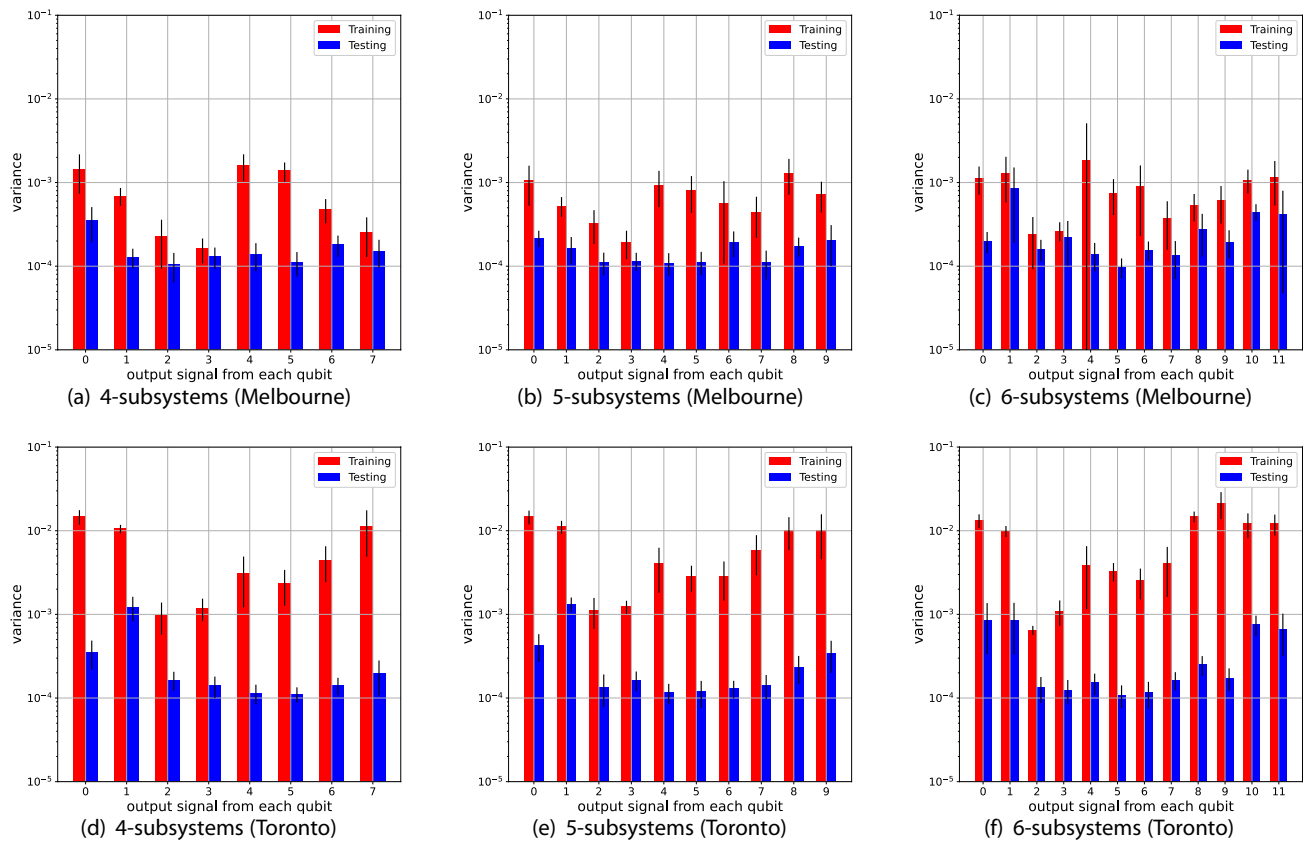

**Figure 7.** Variance of the output signal from each qubit over training (red) and testing (blue) phases. The number on the horizontal axis indicates the label of the qubit from which the signal is measured. The upper three panels (**a**–**c**) and lower three panels (**d**–**f**) show the results for the Melbourne and Toronto device, respectively.

|  | NARMA2 | NARMA5 | NARMA10 |
|---|---|---|---|
| **(a) Training phase** | | | |
| Mean | 0.193 | 0.178 | 0.192 |
| Variance | $1.71 \times 10^{-6}$ | $1.61 \times 10^{-4}$ | $1.74 \times 10^{-4}$ |
| **(b) Testing phase** | | | |
| Mean | 0.193 | 0.182 | 0.197 |
| Variance | $9.14 \times 10^{-7}$ | $4.35 \times 10^{-5}$ | $6.46 \times 10^{-4}$ |

**Table 2.** Mean and Variance of the target output sequence for (a) training and (b) testing phase.

The above-mentioned difference in the stationarity may be attributed to the non-identical noise characteristics resulting from the layout of the device; that is, a device with qubit configuration that is susceptible to noise is more likely to show faster convergence of the dynamics resulting in a more stationary output signal over the training and testing phase. Presumably the Melbourne device suffers from more serious and complicated noise than the Toronto one, due to the difference in the structure design of the device. It is reported in Ref[56] that the CNOT gate error rate for a device with a dense square lattice structure like the Melbourne device is greater than that of one with a sparse hexagonal structure like the Toronto device. Hence, our results would indicate that noise in the Melbourne device is more suitable for the NARMA tasks in the sense of the stationarity over the training and testing period. From this standpoint, we can also understand another device-dependent feature regarding the relation between the performance and the system size; the powerful noise on the Melbourne device results in the stationary output signals that are though lacking in the higher order time series. Therefore, the extra output signals coming from the added system may contribute little to the performance improvement. On the other hand, output signals generated from the Toronto device have higher-order time series due to the moderate noise, which as a result improves the performance as we increase the system size. In this way, our scheme shows the different performance depending on the hardware, even if the operations (i.e. the type of quantum circuits) are the same. Therefore, it is critical to investigate what kind of noise in quantum hardware and structure of quantum processors are preferable for our scheme, which will be left to a future work.

Furthermore we compare our QR system to the standard ESN, as a typical machine learning model. The ESN used here is the RC model given by Eqs. (1) and (2), where $h(x_t) = x_t$ and the function $f$ is expressed as

$$f(x_{t-1}, u_t) = g(W^T x_{t-1} + W_{in}^T u_t),$$

with the hyperbolic tangent function $g$ and the randomly initialized weighting matrices $W_{in}$ and $W$. Specifically, with the number of internal nodes $N_{ESN}$ (i.e., the state vector $x_t \in \mathbb{R}^{N_{ESN}}$), $W_{in}$ is a $N_{ESN}$ dimensional random binary vector, and $W$ is a $N_{ESN} \times N_{ESN}$ matrix whose element is drawn from a standard normal distribution. In general, the performance of the ESN depends on not only $N_{ESN}$ but also the spectral radius[57] of $W$. Hence, we investigate the performance of the ESN over a range of values for these two hyper-parameters; for $N_{ESN} = 2, 5, 10, 20, 50$ and the spectral radius ranging from 0.01 to 1 in increments of 0.01, we performed 100 trials with different $W_{in}$ and $W$ for each experimental setting. Then, for comparison with our QR models, we calculate *the global average of NMSE* and *the global minimum of NMSE* introduced in[17]. The former is the NMSE averaged over all trials for the entire settings of the spectral radius with a fixed number of internal nodes $N_{ESN}$, and can be understood as the expected performance of an arbitrary ESN with $N_{ESN}$ internal nodes. The latter is the minimum NMSE averaged over 100 trials for all the settings of the spectral radius with a fixed $N_{ESN}$ and interpreted as an optimal ESN with $N_{ESN}$ internal nodes. The results for the numerical experiments are summarized in Supplementary Table S1. We observed that an ESN with a larger $N_{ESN}$ shows the better values for these factors in general, as is expected. With these factors of the ESN, we compare the performance of the ESN model with that of our model. Unfortunately we observed that our QR system on the Melbourne device is comparable to the ESN with several nodes, and that on the Toronto device is worse than the ESN with only a few nodes; the Melbourne device is comparable to an optimal ESN with $N_{ESN} = 2$ ($8.9 \times 10^{-6}, 1.5 \times 10^{-3}, 1.2 \times 10^{-3}$ for NARMA2, NARMA5 and NARMA10, respectively), but inferior to an arbitrary ESN with $N_{ESN} = 5$ ($3.5 \times 10^{-6}, 4.9 \times 10^{-4}, 7.7 \times 10^{-4}$ for NARMA2, NARMA5 and NARMA10, respectively), while the Toronto device is not as good as an arbitrary ESN with $N_{ESN} = 2$ ($1.3 \times 10^{-5}, 1.8 \times 10^{-3}, 1.3 \times 10^{-3}$ for NARMA2, NARMA5 and NARMA10, respectively). See more details of the experimental results in Supplementary Table S1. It is not surprising that the ESN outperforms our QR system, because the QR is not originally designed for these machine learning tasks. Moreover, it seems that our scheme is "over-simplified" for the purpose of seeing the contribution of quantum hardware noise, and thus it has plenty room for improvement by tuning hyper-parameters, such as changing the gate sets and the way to inject the input. Thus, our fully-tuned QR system is a potential candidate for strong physical reservoir computing systems.

**Object classification task.** Next we show the performance of our QR systems for the task of classifying different objects, using the time series data generated from the sensor robot that grabs them. This task is equivalent to identifying the class to which the time series data belongs. Hence QR system aims to separate different types of those inputs.

We use the linear regression technique to train the readout weight $W_{out} \in \mathbb{R}^{N+1} \times \mathbb{R}^K$, where $N$ is the number of observed signals from the reservoir state and $K$ is the number of classes. Recall that an additional dimension for $W_{out}$ (i.e. "+1" in $\mathbb{R}^{N+1}$) comes from the extra bias term. To learn the parameters, we simply solve the following linear equation by the pseudo inverse:

$$(Y_1, \ldots, Y_{N_{train}}) = W_{out}^T (\tilde{X}_1, \ldots, \tilde{X}_{N_{train}}),  \quad (15)$$

where $\tilde{X}_i = \left(\tilde{h}(x_{t_s}^i), \ldots, \tilde{h}(x_{t_e}^i)\right)$ and $Y_i = (\bar{y}^i, \ldots, \bar{y}^i)$. Here $x_t^i$ represents the reservoir state at timestep $t$, and $\bar{y}^i$ the target output for the $i$-th training data with the total number of the training data $N_{train}$. Also $t_s$ and $t_e$ represent the first and last timestep for each data, respectively. The target output is expressed by the one-hot vector (i.e., $[1, 0]^T$ for Object A and $[0, 1]^T$ for object B, for the binary-classification task). As for testing, with the optimized parameter $W_{out}^{opt}$ and $\tilde{X}_{new} = \left(\tilde{h}(x_{t_s}^{new}), \ldots, \tilde{h}(x_{t_e}^{new})\right)$ for the unseen testing data, we compute the following:

$$t_{new} = \text{argmax}\left(\text{mean}_t\left(W_{out}^{opt\,T} \tilde{X}_{new}\right)\right). \quad (16)$$

In fact, $t_{new}$ shows the index of the maximum value in the predicted one-hot vector averaged over the used timesteps (i.e. the period between time $t_s$ and $t_e$), and thus indicates the class of the data (indeed, this is equivalent to the Winner-Takes-All strategy). Note that this learning method was employed for the classification task with the RC model, as shown in[58,59].

In this experiment, the following three objects were used; the first object was a cube made of ABS LEGO blocks weighing 15 grams and measuring 3.2 cm wide (Object A) and the remaining two were a polylactic acid (PLA) cube and sphere made using a 3D printer with equal widths of 3 cm (Object B and object C, respectively). A picture of these objects are shown in Fig. 8a. The sensor data of these objects was obtained by grabbing them using the triboelectric nanogenerator (TENG) sensor and the grabbing robot, illustrated in Fig. 8b,c respectively. The TENG sensor is a pressure sensor which uses an electronegative silicone bubble shaped dome and an electropositive nylon layer as the active materials. Figure 9 shows the raw time series data for each object, where the

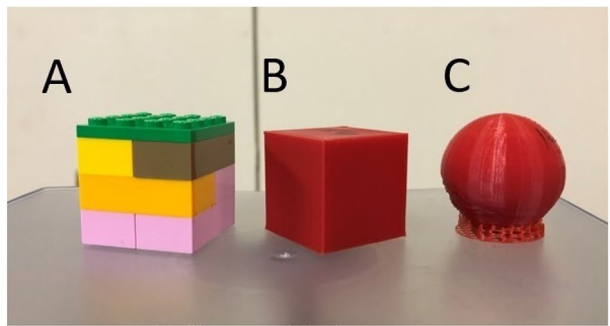
(a) Objects used for the experiments

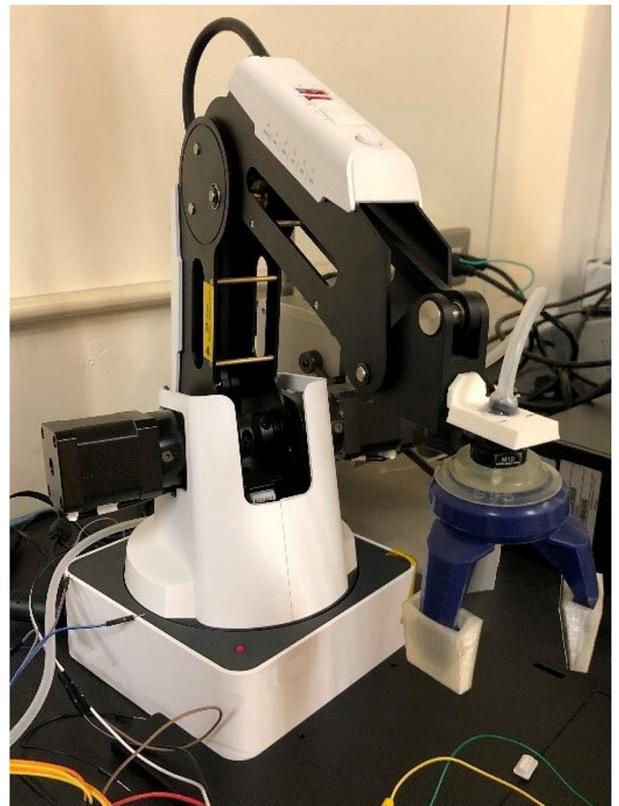

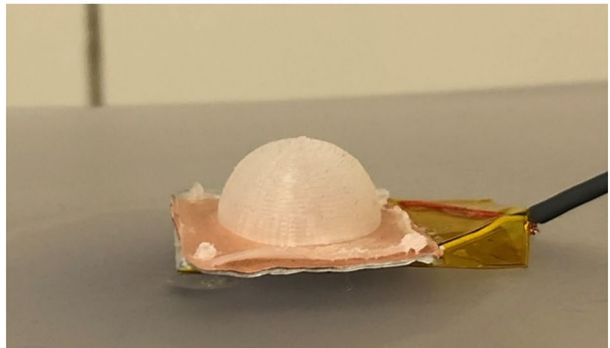
(b) The TENG sensor

(c) The Grab Robot

**Figure 8.** The objects and an instrument used to collect the sensor-data for the classification tasks. Here, the sensor data of three objects (**a**) are obtained using the TENG sensor (**b**), which is manipulated by the grabbing robot (**c**).

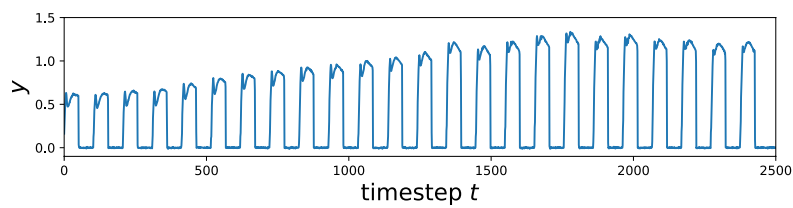
(a) Object A

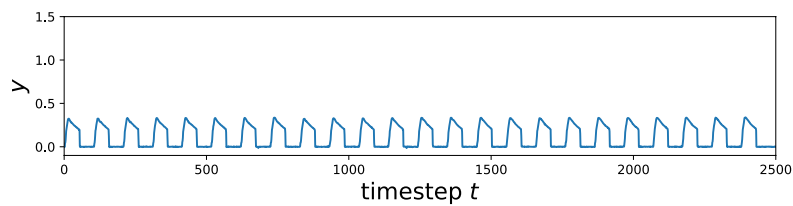
(b) Object B

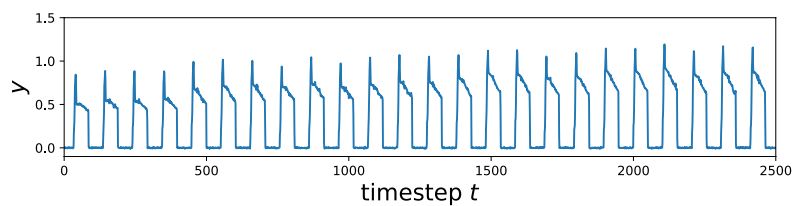
(c) Object C

**Figure 9.** The raw data of each object. (**a**) Object A, (**b**) Object B and (**c**) Object C.

|  | A vs. B | A vs. C | B vs. C | A vs. B vs. C |
|---|---|---|---|---|
| **(a) Our QR system** | | | | |
| Mean | 0.90 | **1.00** | **1.00** | **0.95** |
| Std | 0.20 | 0.00 | 0.00 | 0.11 |
| **(b) Linear regression** | | | | |
| Mean | **1.00** | 0.90 | **1.00** | 0.67 |
| Std | 0.00 | 0.20 | 0.00 | 0.00 |

**Table 3.** Classification accuracy of our QR system and a simple linear regression model for all classification tasks.

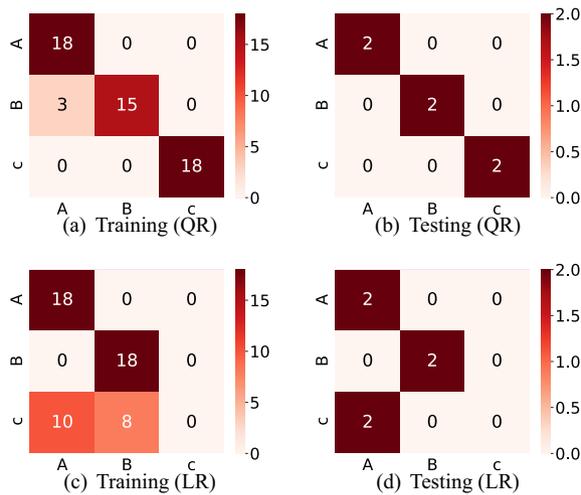

**Figure 10.** The confusion matrices for our QR systems and the linear regression (LR) model in one round of cross-validation. The confusion matrices of QR systems (LR model) for training and testing are shown in (**a**,**c**) and (**b**,**d**), respectively.

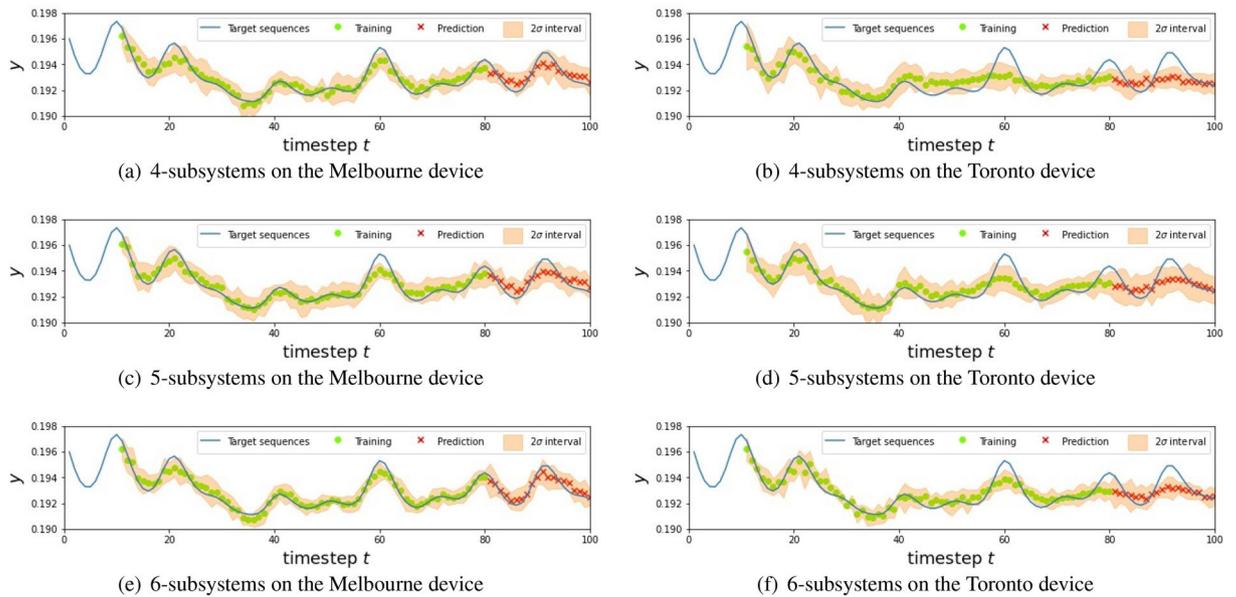

**Figure 11.** The result for NARMA2 using different QR systems. In the figures, the blue line represents the targets, green circles and red crosses are the predictions in the training and testing phase, respectively, and the blur orange regions are $2\sigma$ intervals.

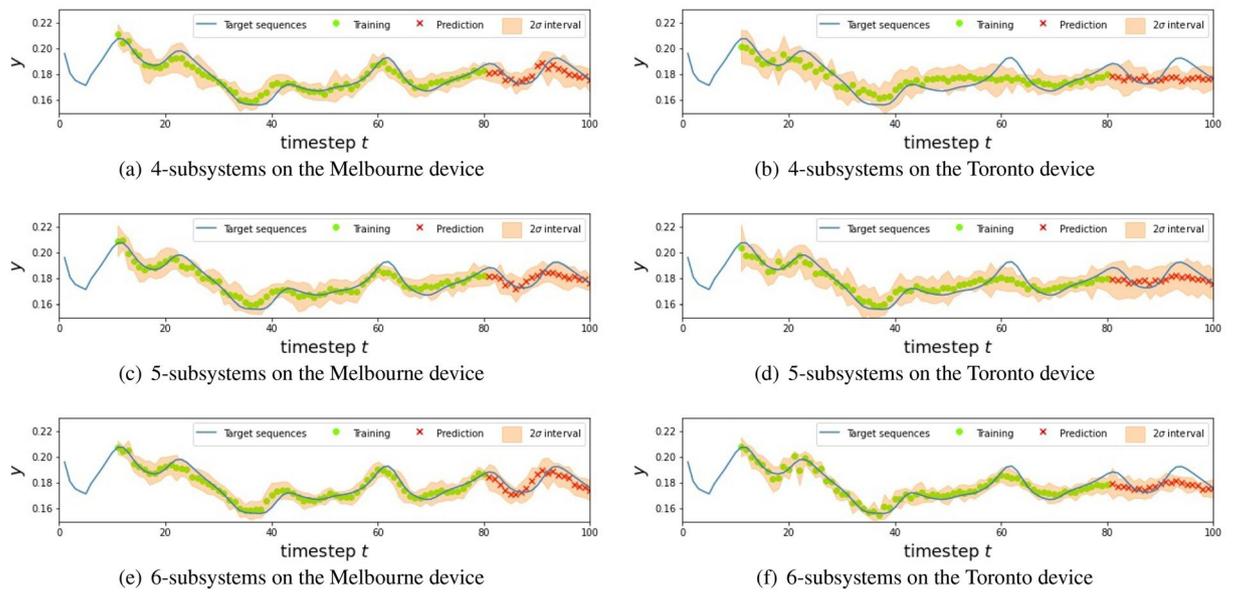

**Figure 12.** The result for NARMA5 using different QR systems. In the figures, the blue line represents the targets, green circles and red crosses are the predictions in the training and testing phase, respectively, and the blur orange regions are $2\sigma$ intervals.

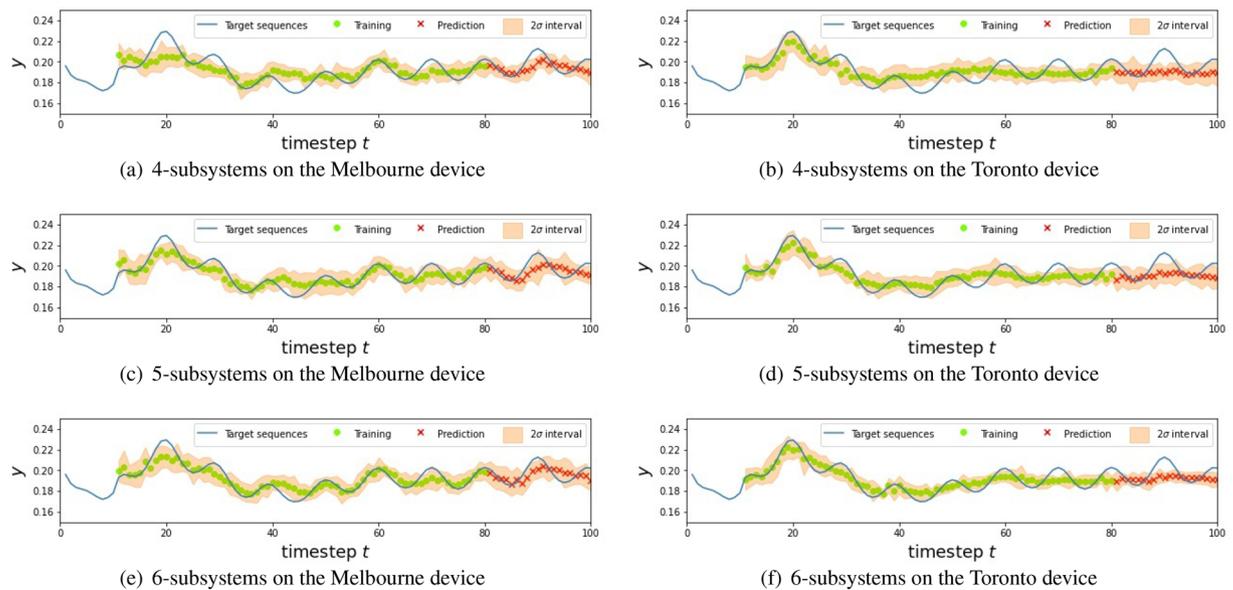

**Figure 13.** The result for NARMA10 using different QR systems. In the figures, the blue line represents the targets, green circles and red crosses are the predictions in the training and testing phase, respectively, and the blur orange regions are $2\sigma$ intervals.

grabbing process was repeated for 25 cycles. We here used 20 pieces of a time-series data obtained in one cycle (90 timesteps) for each object, and pre-processed the data as follows:

$$u_t = u'_{t+1} - u'_t, \tag{17}$$

where $u_t$ and $u'_t$ represent the pre-processed and the raw data at time $t$, respectively. Note that the pre-processing is just taking a finite difference of $u'_t$, and thus its computational cost is negligible.

In this experiment, we consider four classification tasks; three binary classification tasks (A vs. B, A vs. C, and B vs. C) and a three-class classification task (A vs. B vs. C). For all these tasks, we performed 10-fold cross validation to evaluate the performance via the averaged classification accuracy. As for our QR system, we used the 4-subsystems on the Toronto device (subsystems labeled 1 to 4 in Fig. 5) with the same unitary operators in Eq. (9), where we set $a = \pi$. Also, for each data, we discard first 40 timestep for washout and the remaining 49 timesteps are used for the learning, i.e. $t_s = 41$, $t_e = 89$. Note that the experiments were performed during the period from Feb. 22nd to Feb. 23rd in 2021.

The classification accuracies are shown in Table 3, where we also show the performance of a simple linear classification model to compare the results. Here, the simple linear classifier predicts the class of the data at timestep $t$ by $\bar{y}_t = W_{out}^T u_t + b^T$ with $W_{out}, b \in \mathbb{R}^1 \times \mathbb{R}^K$ and predict the testing data according to Eq. (16). We found that our QR system is superior to the linear model for the binary classification task (A vs. C) and the three-class classification, while the linear model is better for the task classifying A and B. Notably, the accuracy of our QR system is almost 0.3 higher than that of the linear model, for the three-class classification task. This result indicates that the QR system can accurately classify three different classes based on inputs, although some classes of which look similar may be more difficult to categorize. As a matter of fact, the confusion matrix of the linear model in one round of cross-validation (Fig. 10) reveals that Object C in the existence of Objects A and B seems difficult to identify, while our QR systems successfully identify them. Moreover, the accuracies for our QR systems are greater than or equal to 0.95 for all the classification tasks, which might be enough comparable to that achieved by some existing effective classifiers such as a logistic classifier. Hence this experiment shows that our proposed QR system also has a potential for executing classification tasks.

## Conclusion

In this work, we propose a QRC scheme that utilizes natural quantum noise; that is, the intrinsic quantum noise, which unavoidably occurs in currently available quantum information devices, is leveraged as a (possibly very rich) computational resource in our QRC framework. We used IBM superconducting quantum processors to experimentally demonstrate the performance of the proposed scheme in two temporal information processing tasks; emulation of the NARMA dynamics and classification of different objects based on sensor-data. As a result, we observed that our proposed scheme outperforms a linear regression or classification model for some NARMA tasks and classification tasks.

There are many remaining works to be examined. In particular, it is important to improve the performance and the processing speed. To improve the performance, we need to somewhat identify the underlying mechanisms of quantum noise on the device under operation, from the perspective of controllability. In fact, recently we found several quantum system identification methods, which aim to elucidate the complex quantum noises such as crosstalk[51,52]. These analyses will help us design a quantum reservoir system with suitable configuration of qubits as well as a unitary gate for the subsystems. Reducing the processing speed of our QR scheme is also important. Processing speed is one of the advantages of physical RC framework compared to other machine learning architectures[20]. In our experimental setting, we have to run the quantum circuits repeatedly to average the stochastic outcomes and obtain the output signals. That is, for the total timesteps $L$ and the number of required measurement $S$, we need to run the quantum circuit a total of $LS$ times, which is time consuming. However, thanks to the recent rapid advance of the quantum hardware, we can now perform mid-circuit measurements for IBM superconducting processors[60], which generates the output signal with only $S$ running the circuit, meaning that our RC method is scalable.

Finally note that the proposed QRC scheme is not limited to NISQ. Even in the era of future fault-tolerant quantum computing devices, it is easy to introduce noise by simply relaxing the control level for qubits. Therefore the degree of noise level introduced as well as the structure of noise dynamics will be a central topic to be explored in the future. Also, in addition to the important direction to explore relevant applications Ref[51,62], an intriguing open problem is to characterize "quantum-reservoir oriented" processors which are specialized for our QRC scheme; examples of such processors are a system with qubits configuration suitable for machine learning tasks and a system whose hyper-parameters such as noise level are tunable.

## References
 1. Baldi, P., Brunak, S. & Bach, F. *Bioinformatics: The Machine Learning Approach* (MIT Press, 2001).
 2. He, K., Zhang, X., Ren, S. & Sun, J. Deep residual learning for image recognition. In *Proceedings of the IEEE Conference on Computer Vision and Pattern Recognition* 770–778 (2016).
 3. Ronneberger, O., Fischer, P. & Brox, T. U-net: Convolutional networks for biomedical image segmentation. In *International Conference on Medical Image Computing and Computer-assisted Intervention* 234–241 (Springer, 2015).
 4. Dixon, M. F., Halperin, I. & Bilokon, P. *Machine Learning in Finance* (Springer, 2020).
 5. Mullainathan, S. & Spiess, J. Machine learning: An applied econometric approach. *J. Econ. Perspect.* **31**, 87–106 (2017).
 6. Greydanus, S., Dzamba, M. & Yosinski, J. Hamiltonian neural networks. arXiv preprint arXiv:1906.01563 (2019).
 7. Hermann, J., Schätzle, Z. & Noé, F. Deep-neural-network solution of the electronic Schrödinger equation. *Nat. Chem.* **12**, 891–897 (2020).
 8. Young, T., Hazarika, D., Poria, S. & Cambria, E. Recent trends in deep learning based natural language processing. *IEEE Comput. Intell. Mag.* **13**, 55–75 (2018).
 9. Schaal, S. & Atkeson, C. G. Learning control in robotics. *IEEE Robot. Autom. Mag.* **17**, 20–29 (2010).
10. Mandic, D. & Chambers, J. *Recurrent Neural Networks for Prediction: Learning Algorithms, Architectures and Stability* (Wiley, 2001).
11. Jaeger, H. & Haas, H. Harnessing nonlinearity: Predicting chaotic systems and saving energy in wireless communication.. *Science* **304**, 78–80 (2004).
12. Jaeger, H. The "echo state" approach to analysing and training recurrent neural networks-with an erratum note. *Bonn, Germany* **148**, 13 (2001).
13. Maass, W., Natschläger, T. & Markram, H. Real-time computing without stable states: A new framework for neural computation based on perturbations. *Neural Comput.* **14**, 2531–2560 (2002).
14. Bishop, C. M. *Pattern Recognition and Machine Learning* (Springer, 2006).
15. Schrauwen, B., D'Haene, M., Verstraeten, D. & Van Campenhout, J. Compact hardware liquid state machines on fpga for real-time speech recognition. *Neural Netw.* **21**, 511–523 (2008).
16. Fernando, C. & Sojakka, S. *Pattern Recognition in a Bucket. European Conference on Artificial Life* 588–597 (Springer, 2003).

17. Nakajima, K., Hauser, H., Li, T. & Pfeifer, R. Information processing via physical soft body. *Sci. Rep.* **5**, 1–11 (2015).
18. Caluwaerts, K. *et al.* Design and control of compliant tensegrity robots through simulation and hardware validation. *J. R. Soc. Interface* **11**, 20140520 (2014).
19. Torrejon, J. *et al.* Neuromorphic computing with nanoscale spintronic oscillators. *Nature* **547**, 428–431 (2017).
20. Tanaka, G. *et al.* Recent advances in physical reservoir computing: A review. *Neural Netw.* **115**, 100–123 (2019).
21. Nakajima, K. Physical reservoir computing-an introductory perspective. *Jpn. J. Appl. Phys.* **59**, 060501 (2020).
22. Fujii, K. & Nakajima, K. Harnessing disordered-ensemble quantum dynamics for machine learning. *Phys. Rev. Appl.* **8**, 024030 (2017).
23. Arute, F. *et al.* Quantum supremacy using a programmable superconducting processor. *Nature* **574**, 505–510 (2019).
24. Harrow, A. W. & Montanaro, A. Quantum computational supremacy. *Nature* **549**, 203–209 (2017).
25. Aaronson, S. & Chen, L. Complexity-theoretic foundations of quantum supremacy experiments. arXiv preprint arXiv:1612.05903 (2016).
26. Bremner, M. J., Montanaro, A. & Shepherd, D. J. Average-case complexity versus approximate simulation of commuting quantum computations. *Phys. Rev. Lett.* **117**, 080501 (2016).
27. Chen, J. & Nurdin, H. I. Learning nonlinear input-output maps with dissipative quantum systems. *Quantum Inf. Process.* **18**, 1–36 (2019).
28. Chen, J., Nurdin, H. I. & Yamamoto, N. Temporal information processing on noisy quantum computers. *Phys. Rev. Appl.* **14**, 024065 (2020).
29. Govia, L., Ribeill, G., Rowlands, G., Krovi, H. & Ohki, T. Quantum reservoir computing with a single nonlinear oscillator. *Phys. Rev. Res.* **3**, 013077 (2021).
30. Martínez-Peña, R., Nokkala, J., Giorgi, G. L., Zambrini, R. & Soriano, M. C. Information processing capacity of spin-based quantum reservoir computing systems. *Cognit. Comput.* 1–12 (2020).
31. Nakajima, K., Fujii, K., Negoro, M., Mitarai, K. & Kitagawa, M. Boosting computational power through spatial multiplexing in quantum reservoir computing. *Phys. Rev. Appl.* **11**, 034021 (2019).
32. Kutvonen, A., Fujii, K. & Sagawa, T. Optimizing a quantum reservoir computer for time series prediction. *Sci. Rep.* **10**, 1–7 (2020).
33. Tran, Q. H. & Nakajima, K. Higher-order quantum reservoir computing. arXiv preprint arXiv:2006.08999 (2020).
34. Ghosh, S., Opala, A., Matuszewski, M., Paterek, T. & Liew, T. C. Reconstructing quantum states with quantum reservoir networks. *IEEE Trans. Neural Netw. Learn. Syst.* **32**, 3148–3155 (2020).
35. Tran, Q. H. & Nakajima, K. Learning temporal quantum tomography. arXiv preprint arXiv:2103.13955 (2021).
36. Negoro, M., Mitarai, K., Fujii, K., Nakajima, K. & Kitagawa, M. Machine learning with controllable quantum dynamics of a nuclear spin ensemble in a solid. arXiv preprint arXiv:1806.10910 (2018).
37. Biamonte, J. *et al.* Quantum machine learning. *Nature* **549**, 195–202 (2017).
38. Gyongyosi, L. & Imre, S. A survey on quantum computing technology. *Comput. Sci. Rev.* **31**, 51–71 (2019).
39. Havlíček, V. *et al.* Supervised learning with quantum-enhanced feature spaces. *Nature* **567**, 209–212 (2019).
40. Mitarai, K., Negoro, M., Kitagawa, M. & Fujii, K. Quantum circuit learning. *Phys. Rev. A* **98**, 032309 (2018).
41. Gyongyosi, L. & Imre, S. Training optimization for gate-model quantum neural networks. *Sci. Rep.* **9**, 1–19 (2019).
42. Preskill, J. Quantum computing in the nisq era and beyond. *Quantum* **2**, 79 (2018).
43. Jaeger, H. Adaptive nonlinear system identification with echo state networks. *Adv. Neural Inf. Process. Syst.* **15**, 609–616 (2002).
44. Verstraeten, D., Schrauwen, B., Stroobandt, D. & Van Campenhout, J. Isolated word recognition with the liquid state machine: A case study. *Inf. Process. Lett.* **95**, 521–528 (2005).
45. Buehner, M. & Young, P. A tighter bound for the echo state property. *IEEE Trans. Neural Netw.* **17**, 820–824 (2006).
46. Meurer, T., Graichen, K. & Gilles, E.-D. *Control and Observer Design for Nonlinear Finite and Infinite Dimensional Systems* Vol. 322 (Springer Science & Business Media, 2005).
47. Nielsen, M. A. & Chuang, I. *Quantum Computation and Quantum Information* (Springer, 2002).
48. Altafini, C. & Ticozzi, F. Modeling and control of quantum systems: An introduction. *IEEE Trans. Autom. Control* **57**, 1898–1917 (2012).
49. Schirmer, S. & Wang, X. Stabilizing open quantum systems by Markovian reservoir engineering. *Phys. Rev. A* **81**, 062306 (2010).
50. Morris, J., Pollock, F. A. & Modi, K. Non-markovian memory in ibmqx4. arXiv preprint arXiv:1902.07980 (2019).
51. Sarovar, M. *et al.* Detecting crosstalk errors in quantum information processors. *Quantum* **4**, 321 (2020).
52. Winick, A., Wallman, J. J. & Emerson, J. Simulating and mitigating crosstalk. arXiv preprint arXiv:2006.09596 (2020).
53. Hochreiter, S. & Schmidhuber, J. Long short-term memory. *Neural Comput.* **9**, 1735–1780 (1997).
54. Kubota, T., Nakajima, K. & Takahashi, H. Dynamical anatomy of narma10 benchmark task. arXiv preprint arXiv:1906.04608 (2019).
55. Atiya, A. F. & Parlos, A. G. New results on recurrent network training: Unifying the algorithms and accelerating convergence. *IEEE Trans. Neural Netw.* **11**, 697–709 (2000).
56. The ibm quantum heavy hex lattice. https://research.ibm.com/blog/heavy-hex-lattice#fn-4. (Accessed: 2021-10-30).
57. Lukoševičius, M. A practical guide to applying echo state networks. In *Neural networks: Tricks of the Trade* 659–686 (Springer, 2012).
58. Araujo, F. A. *et al.* Role of non-linear data processing on speech recognition task in the framework of reservoir computing. *Sci. Rep.* **10**, 1–11 (2020).
59. Larger, L. *et al.* High-speed photonic reservoir computing using a time-delay-based architecture: Million words per second classification. *Phys. Rev. X* **7**, 011015 (2017).
60. Mid-circuit measurements tutorial. https://quantum-computing.ibm.com/lab/docs/iql/manage/systems/midcircuit-measurement/. (Accessed: 2021-05-11).
61. Alexeev, Y. *et al.* Quantum computer systems for scientific discovery. *PRX Quantum* **2**, 017001 (2021).
62. Awschalom, D. *et al.* Development of quantum interconnects (quics) for next-generation information technologies. *PRX Quantum* **2**, 017002 (2021).

### Acknowledgements
This work was supported by MEXT Quantum Leap Flagship Program Grant Numbers JPMXS0118067285 and JPMXS0120319794, and also JSPS KAKENHI Grant Number 20H05966. Y.S. and N.Y. acknowledge the discussion with Kohei Nakajima and Tran Hoan.

### Author contributions
Y.S. Q.G., and N.Y. designed the research. Y.S. conducted the experiments on the quantum devices. K.C.P. conducted the experiments to obtain the sensor-data. Y.S and N.Y. wrote the manuscript. All authors analyzed the results and contributed to the continuous improvement of the manuscript.

### Competing interests
The authors declare no competing interests.

### Additional information
**Supplementary Information** The online version contains supplementary material available at https://doi.org/10.1038/s41598-022-05061-w.

**Correspondence** and requests for materials should be addressed to Y.S.

**Reprints and permissions information** is available at www.nature.com/reprints.

**Publisher's note**  Springer Nature remains neutral with regard to jurisdictional claims in published maps and institutional affiliations.

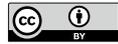 **Open Access**  This article is licensed under a Creative Commons Attribution 4.0 International License, which permits use, sharing, adaptation, distribution and reproduction in any medium or format, as long as you give appropriate credit to the original author(s) and the source, provide a link to the Creative Commons licence, and indicate if changes were made. The images or other third party material in this article are included in the article's Creative Commons licence, unless indicated otherwise in a credit line to the material. If material is not included in the article's Creative Commons licence and your intended use is not permitted by statutory regulation or exceeds the permitted use, you will need to obtain permission directly from the copyright holder. To view a copy of this licence, visit http://creativecommons.org/licenses/by/4.0/.

© The Author(s) 2022